\documentclass[12pt]{iopart}
\usepackage{graphicx}
\usepackage{cite}
\pdfoutput=1 % if your are submitting a pdflatex (i.e. if you have
\begin{document}

\title{Tagging $b$ quarks at extreme energies without tracks}
%\title{\boldmath A title with some math: $x=1$}

%% %simple case: 2 authors, same institution
\author{B. Todd Huffman, Charles Jackson, and Jeff Tseng}
%JHEP control sequence
%\affiliation{Oxford University dept. of Particle Physics,\\Oxford, United Kingdom} 
\address{Particle Physics, Oxford University \\ 
	Keble Road \\ Oxford  OX1 3RH \\
	United Kingdom}
\ead{todd.huffman@physics.ox.ac.uk}
\vspace{10pt}
\begin{indented}
\item[] 14 June 2016 %\today
\end{indented}

\begin{abstract}
We describe a new hit-based $b$-tagging technique for high energy jets and 
study its performance with a {\sc Geant4}-based simulation. 
The technique uses the fact that at sufficiently high energy a $B$~meson or baryon can live long
enough to traverse the inner layers of pixel detectors such as those in the 
ATLAS, ALICE, or CMS experiments prior to decay. 
By first defining a ``jet'' via the calorimeter, and then
counting hits within that jet between pixel layers at increasing radii, 
we show it is possible to identify jets that contain $b$-quarks by detecting a jump in the 
number of hits without tracking requirements. 
We show that the technique maintains fiducial efficiency at TeV scale 
$B$ hadron energies, far beyond the range of existing algorithms, 
and improves upon conventional $b$-taggers.
\end{abstract}

\section{Introduction}
\label{sec:intro}

Many of the most exciting searches for new physics beyond the Standard Model,
as well as further studies of the Standard Model itself, benefit from being
able to identify high-energy jets containing $b$ quarks (``$b$-jets'').
Examples include Higgs pair production and decay via
$HH\rightarrow b\overline{b}b\overline{b}$, sensitive to Higgs trilinear
couplings~\cite{Behr:2015oqq}; graviton and radion decays to heavy fermions
and bosons in warped extra dimension models~\cite{Gouzevitch:2013qca}; third-generation
superpartners in supersymmetry~\cite{Alwall:2008ag}; and indeed any new
physics with preferential couplings to heavy Standard Model particles or
third-generation fermions in particular.

One of the most distinctive features of a $b$-jet is the relatively long life
(on the order of 1.5~ps) of the $B$ hadron, resulting in charged particle
tracks displaced from the primary interaction vertex.  For this reason, almost
all modern collider-based particle physics experiments deploy several layers of
high-granularity silicon detectors near the interaction point. Algorithms
for distinguishing $b$-jets from jets originating from lighter quarks rely on
reconstructed high-resolution tracks in these finely grained subsystems.

However, with increasingly stringent limits placed on the energy scale for new physics, 
distinguishing displaced tracks within increasingly energetic jets
becomes simultaneously more important and more challenging.  Two effects in
particular make $b$-tagging in TeV-scale jets difficult:  First,
more tracks are collimated into a small angle, resulting in a higher hit
density and a more ambiguous association of hits with tracks.
A single mis-assignment can steer a track off-course and produce an
erroneous impact parameter.  Second, at extreme energies, an increasing
fraction of $B$ hadrons will decay after crossing the innermost layers of the
silicon detector:  in the best case scenario, this situation merely reduces the
number of hits available for reconstruction and thus degrades the impact
parameter resolution of the track.  A worse scenario is that the track picks up
a spurious hit in the densely populated inner layer.

Results on conventional $b$-tagger efficiencies from the LHC experiments
are limited to momenta transverse to the beam ($p_T$) below 
600~GeV\cite{ATL-PHYS-PUB-2015-022}, and show falling tagging efficiency
beyond approximately 150 GeV.  Nevertheless, the ATLAS experiment has
measured the invariant mass spectrum of $b$ hadron enriched jets
at $\sqrt{s}=13\;{\rm TeV}$ out to 5~TeV\cite{Aaboud:2016nbq}, illustrating both the
importance and challenges of the highly boosted regime. 
It is also worth noting the development of other $b$-tagging algorithms
dedicated to this regime, including initial studies of muon-based
tagging~\cite{Pedersen:2015knf} using the {\sc Delphes}
parametric detector simulation~\cite{deFavereau:2013fsa}.

This article investigates a new method which, by relying only on the hits rather 
than the reconstructed tracks, better maintains its efficiency 
at extreme energies, 
by which we mean energies of at least 600~GeV, above which conventional 
hadronic $b$-tagging performance degrades. Section~\ref{sec:algorithm} 
describes this new
method, which we call ``multiplicity jump'' $b$-tagging.
Section~\ref{sec:simulation} outlines the simulation used to test the method,
with the results given in Section~\ref{sec:Perform}.
Section~\ref{sec:conclusion} then concludes and describes 
prospects for further study.

\section{The ``multiplicity jump'' $b$-tagger}
\label{sec:algorithm}

\begin{figure}[thbp]
\centering 
\includegraphics[height=.6\textwidth]{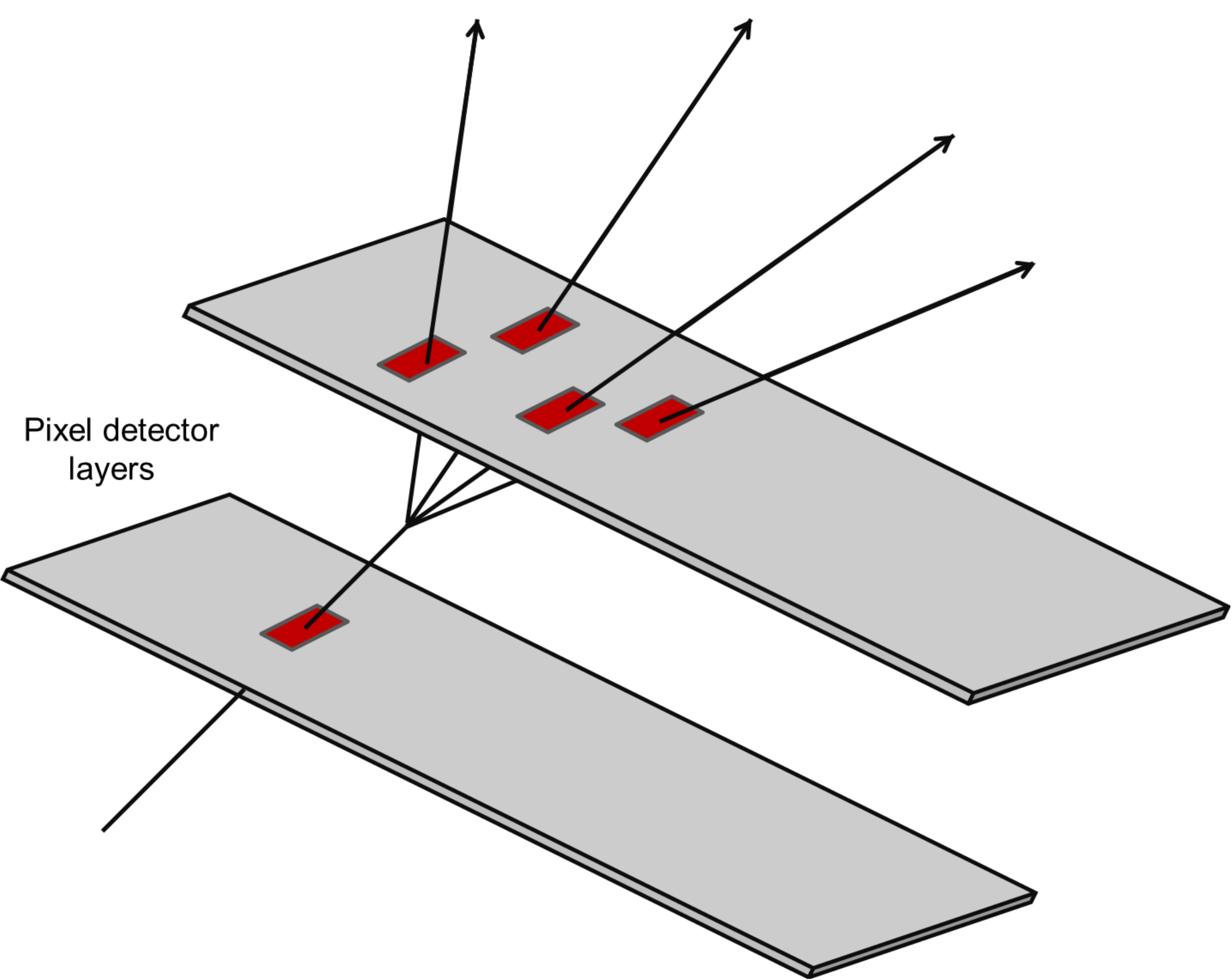}
\caption{The ``multiplicity jump'' tagger works when a particle with a large 
lorentz boost decays between two layers of pixel detectors. 
Shown here schematically is a particle traversing a pixel layer from the 
lower left and decaying before the next layer, causing multiple hits to appear. 
For this tag to be most effective the particle should decay into many daughter particles. 
$B$ hadrons have this desirable property.}
\label{fig:BotOxJump}
\end{figure}

As mentioned above, almost all modern collider-based experiments deploy
high-granularity detectors near the interaction point, in particular so-called
``pixel'' detectors.  For the present discussion, we work in a cylindrical
coordinate system in which the origin is located at the nominal
interaction point, $z$ is measured along the beamline, and $r$ and $\phi$ are
the radius and azimuthal angle in the plane transverse to the beam.  The angle
$\theta$ is measured relative to the beam, and pseudorapidity is defined as
$\eta\equiv -\ln\tan(\theta/2)$.  A pixel detector, such as that used in the
ATLAS experiment, is envisaged as several cylindrical layers of silicon sensors placed at
increasing radii from the interaction point\cite{AtlasMaterial}. Silicon pixel sensors are 
similar to the pixels within a digital camera and consist of many hundreds of thousands
of individual sensors (the pixel channels) which can each register a signal 
when a charged particle passes through them.
This is recorded as a ``hit channel'' or just a ``hit''. 

The multiplicity jump algorithm seeks to tag $B$ hadron decays between the
pixel layers as shown schematically in figure~\ref{fig:BotOxJump}.  
Such decays usually increase the number of charged particles
traversing subsequent detector 
layers, and thus should be observable as an increase in
the number of hits in a small angular region, defined as the area within $\Delta
R\equiv\sqrt{\Delta \eta^2+\Delta\phi^2}<0.04$ relative to some pre-defined jet
axis. The small radius is close to the expected angular spread $2m/p_T$ of 
the decay products of a $B$ hadron
with momentum transverse to the beam, $p_T$, in excess of approximately
300~GeV.\footnote{$\Delta R < 0.4$ and $\Delta R < 0.1$ were also explored 
but did not achieve better separation.}
Such a cone is too narrow for most calorimeters, but easily spans
numerous pixels.  The number of hits $N_j$ in pixel layer $j$, counting up from
the innermost layer, is calculated by counting the hits within the angular
region.  The (relative) multiplicity jump $f_j$ at layer $j$ is then defined to
be
\begin{equation}
f_j=\frac{N_{j+1}-N_j}{N_j}=\frac{\Delta N_j}{N_j}.
\label{eq:reljump}
\end{equation}
For example, $f_j=1$ indicates that there are twice as many hits in layer $j+1$
than in layer $j$.  A jet is tagged as a $b$-jet if $f_j$ exceeds a value $F$
for any pair of layers $j$ and~$j+1$.  It is worth noting that sequential charm 
decay can also generate a positive multiplicity jump.

An absolute multiplicity jump was also considered, 
\begin{equation}
\Delta N_j = N_{j+1} - N_j
\end{equation}
but discarded due to the effect of showering, which is
expected to increase the number of hits in proportion to the number of
particles (and therefore hits).  As a result, $\Delta N_j$ is expected to
increase with $j$.  On the other hand, showering should add a mostly
layer-independent offset to $f_j$.  Setting the tag threshold $F$ appropriately
should then reduce the algorithm's sensitivity to showering.
  
The idea of using a multiplicity jump as a method for tagging $b$-jets is not
new.  Early bottom and charm fixed target experiments attempted a similar
method using multiple planes of scintillators or Cherenkov radiation 
detectors\cite{JumpMult}.  The integrated signal from an
upstream scintillator was compared to that from a matched downstream
scintillator, and a ``jump'' in signal provided the heavy flavour tag.  Such
methods faced challenges due to large fluctuations in the energy deposited by
relativistic particles.  The present method, on the other hand,
relies on the vastly increased granularity of pixel detectors and the relative,
rather than absolute, multiplicity jump.
  
\section{Simulation}
\label{sec:simulation}

\begin{figure}[tbhp]
\centering 
\includegraphics[height=.33\textwidth]{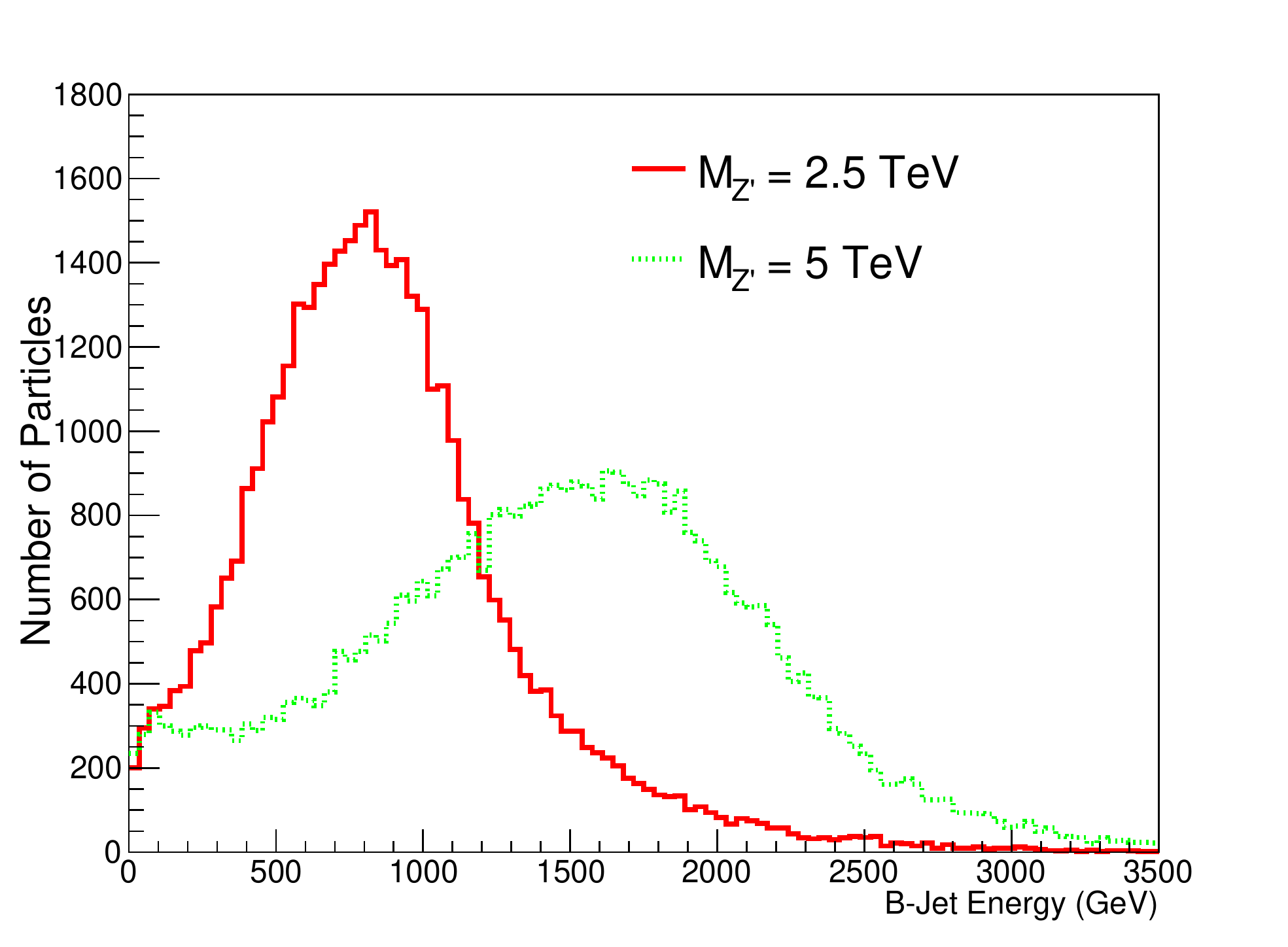}
\includegraphics[height=.33\textwidth]{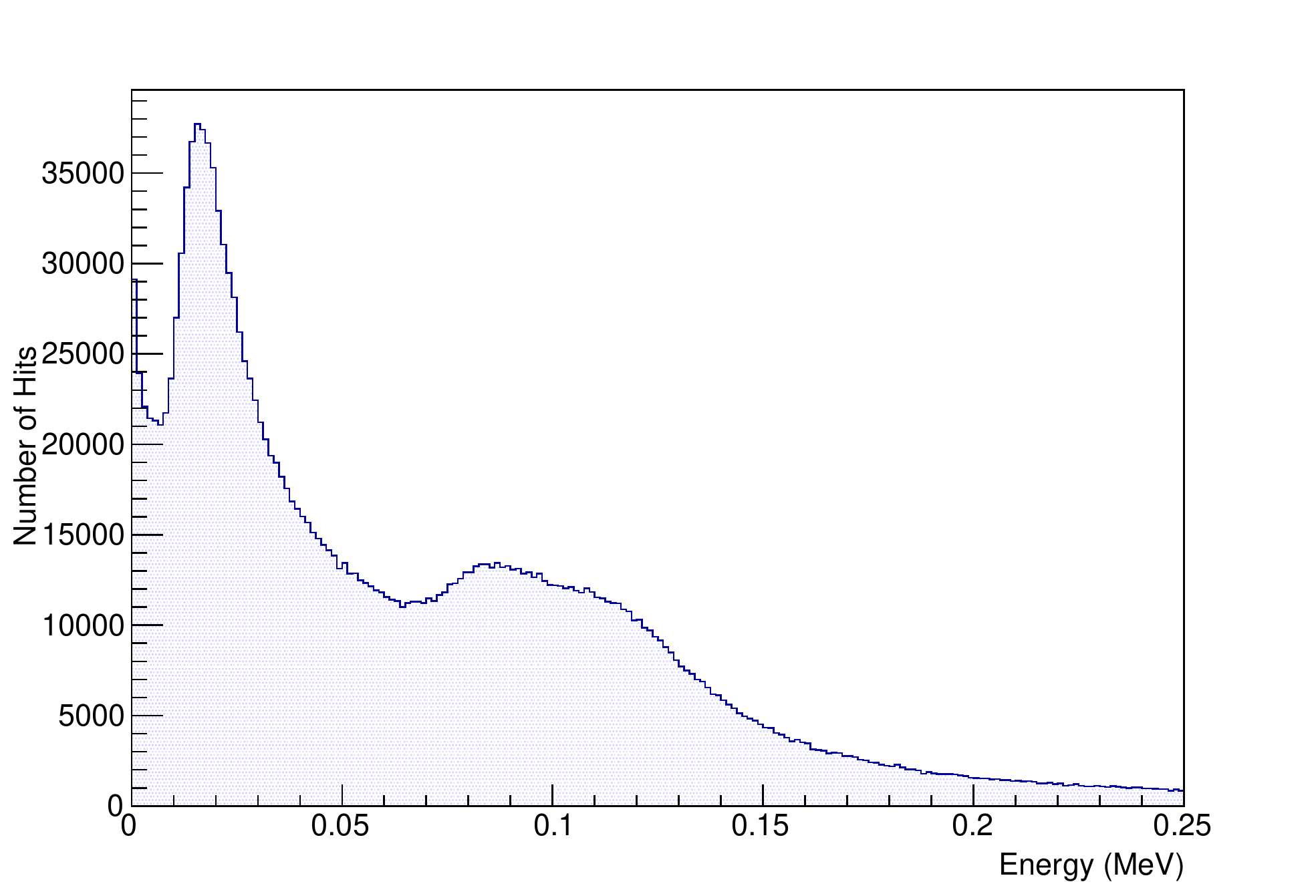}
\caption{Left:  energy distributions of $b$ jets in simulated samples with $Z'$
masses of 2.5 and 5~TeV where the jet has been clustered using the anti-$kt$ 
algorithm from {\sc FastJet} with $R = 0.2$.
Right:  energy deposition in individual pixel volumes, for the sample with
2.5~TeV $Z'$ mass.  Zero energy deposition has been suppressed.}
\label{fig:eb}
\end{figure}

The new method was tested in a simulation based on {\sc Geant4} (version 10.0)
in order to model particle interactions and showering in a 
detector\cite{Agostinelli:2002hh}\cite{Allison:2006ve}.  
{\sc Pythia} version 8.209\cite{Pythia8}, with the default Monash 2013 
tune\cite{Skands:2014pea}, was used to simulate
$pp$ collisions with center-of-mass energy $\sqrt{s}=13\;{\rm TeV}$.
High-energy $b$-jets and those with lighter quarks were generated by creating
$Z'$ bosons with masses of 2.5 and 5~TeV.  The $Z'$ bosons were forced to decay to
$q\overline{q}$ pairs, where $q$ is any quark but $t$, and hadronization and
fragmentation handled by {\sc Pythia}.  Initial and final state radiation
resulted in jets with a range of momenta mostly below $M_{Z'}/2$.
The $B$ hadron takes most of the jet energy, with
the most likely energy fraction around 85\%, independent of the 
initial parton energy. The $b$-jet 
energy distribution is shown in Figure~\ref{fig:eb}(left).
The $B$ hadron was observed to take most of the jet energy in a 
manner quantitatively similar to~\cite{Peterson:1982ak}.
Decays of $B$ hadrons
were then simulated using {\sc EvtGen} version 1.4.0, with bremsstrahlung
handled by {\sc Photos} version 3.52 and any $\tau$ decays by {\sc Tauola}
version 1.0.7\cite{Lange:2001uf}.

%no special tune.  softQCD off.  13 TeV.
% hepMC 2.06.08

A simplified detector geometry, loosely based on the four-layer ATLAS pixel barrel
system, was used to model the detector response.  The active pixel layers, with radii
25.7, 50.5, 88.5, and 122.5~mm, were encased within a volume of air and inside
a uniform 2~T magnetic field pointing in the positive $z$ direction.  Each
barrel was 1.3~m long (the innermost layer, the ``Insertable $B$ Layer'' or
IBL, of the ATLAS pixel system is actually slightly shorter\cite{AtlasTDR}).  The pixel
sensors were $300\;{\rm \mu m}$ thick, with a $50\;{\rm \mu m}$ pitch in the
$\phi$ direction, and a $400\;{\rm \mu m}$ length in the $z$ direction
($800\;{\rm \mu m}$ in our innermost layer in order to test the effect of 
varying granularity).  These idealized pixels were simulated as pure
silicon slabs without gaps.

It is worth noting that the geometry largely determines the energy range in which
the multiplicity jump algorithm works best:  approximately 300~GeV is required
for the average $B$ hadron flight distance to reach the innermost layer.
Beyond approximately 1.4~TeV, the average flight distance reaches beyond the
outermost layer.

In order to model inactive material such as the support structure, mountings,
cooling pipes, and electronics, we added further cylinders of silicon to the {\sc
Geant4} model, located just outside each cylinder of sensitive pixels, so as to
bring the total simulated material up to an equivalent of 2.5\% of radiation
length per layer. In addition a silicon cylinder half as thick was added just inside the
outermost active layer of pixels.

% no beampipe

Figure~\ref{fig:eb}(right) shows the non-zero energy deposition in all pixels for 5000 events
modelled in {\sc Geant4}.  The broad peak around 0.1~MeV corresponds to a
minimum ionizing particle at roughly normal incidence, the broadness is partly a
result of the long tail in the charged particle energy loss distribution.  
The sharp peak just above 0.02~MeV originates from low energy particles 
curling within the magnetic field, traversing the $50\;{\rm \mu m}$ width 
of the pixel.  The peak near zero energy corresponds to low energy products 
of interactions within pixels propagating partially into neighboring pixels.  
Since we are concerned with particles which are energetic enough to be mostly
normally incident, we impose a threshold of 0.05~MeV (well above the ATLAS
threshold of 0.011~MeV\cite{AtlasTDR}) before we register the pixel as having 
been ``hit''. No attempt has been made to form either clusters nor tracks 
from individual pixel hits.

Since the simulation does not extend to calorimeters, we cluster stable
generated particles (excluding neutrinos) using the {\sc FastJet} 
(version 3.1.3)\cite{FastJet} implementation of the ``anti-$k_T$'' 
sequential recombination algorithm\cite{antikt} with $R=0.2$ (the ATLAS 
hadronic calorimeter granularity is approximately $0.1\times
0.1$ in $\phi\times\eta$).  The jet's axis is used to define the 
angular region in the multiplicity jump algorithm.

The sample of $b$ jets is defined by finding the highest energy ground state
$B$ hadron within $\Delta R<0.2$ of the jet axis. After $b$ jets are so identified,
a similar search is performed to identify charm jets. All other jets are considered
``light quark'' jets (or ``uds'' jets). The two
highest energy $b$-jets are then used to test the efficiency of the
multiplicity jump algorithm. It should be noted that using these criteria, 
13\% of $b$ jets have the $B$ hadron within $\Delta R <0.2$ but outside 
of $\Delta R<0.04$. Such $B$ hadrons contribute to an inefficiency in the algorithm. 

\section{Performance}
 \label{sec:Perform}

In order to measure the algorithm's performance in our simulation, we define
an efficiency $\epsilon_b$ for $b$ jets in a fiducial region as the number
of tagged $b$ jets divided by the number of jets in which the matched $B$
hadron decays in the fiducial region.  The fiducial region is defined in
terms of the inner and outer pixel layers being investigated;
in other words, the $\epsilon_b$ reflects the probability that, if a $B$
hadron decays between two pixel layers, it will be tagged by the algorithm.

We note that the fiducial volume as defined does not capture all those
$b$ jets which could be tagged in principle.  For instance, a $B$ hadron
which decays just before the inner layer could leave one hit in that layer,
and a multiplicity jump in the next.  Likewise, a $B$ hadron which decays
just before the outer layer could be impossible to recognize because its
hits are merged into one or a few pixels.  The simple fiducial volume,
however, is sufficient for the present purpose of examining the
algorithm's basic behavior.

\begin{figure}[thbp]
\centering 
\includegraphics[width=.99\textwidth]{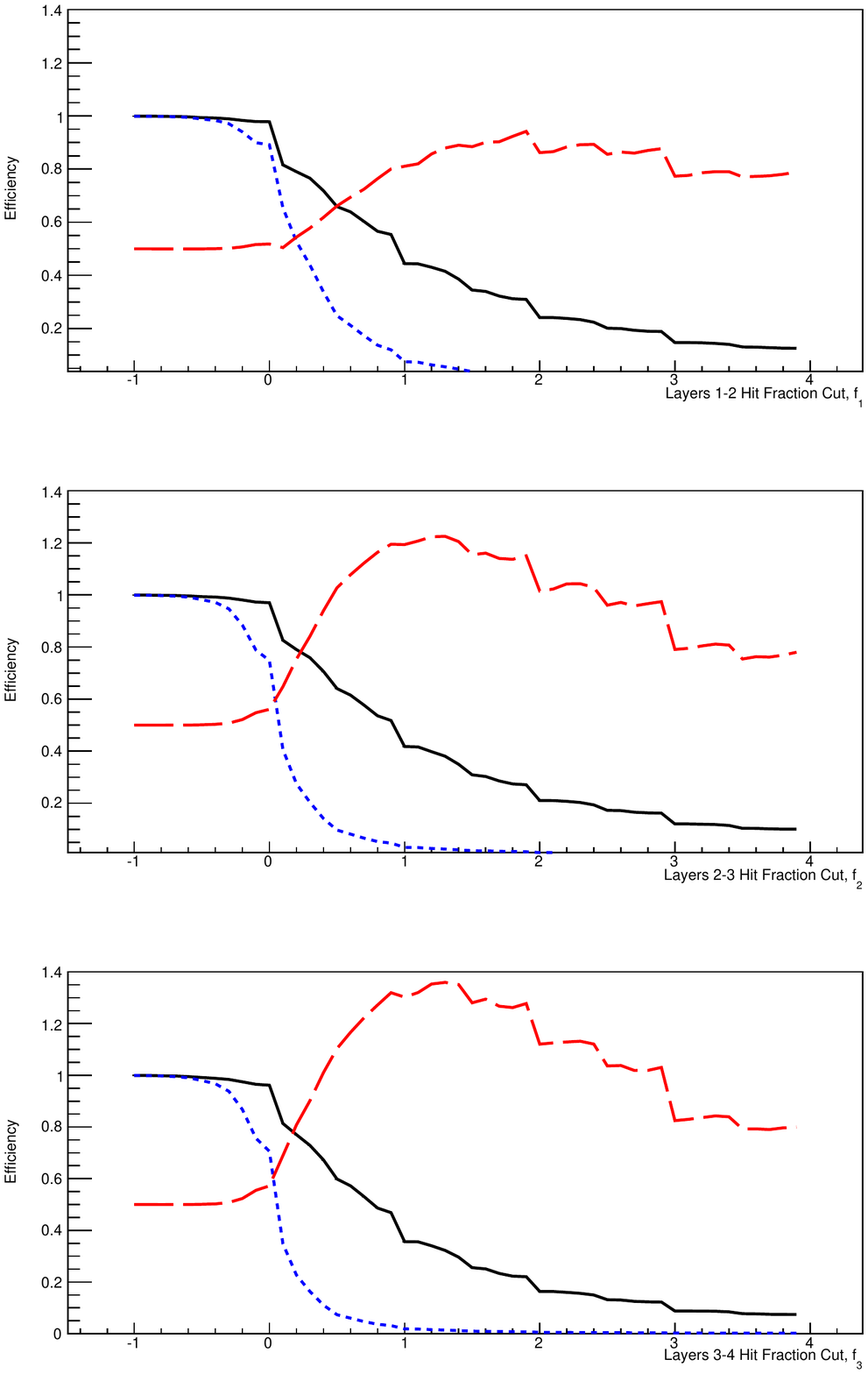}
\caption{$\epsilon_b$(solid line) and $\epsilon_q$(short-dashed line) for fiducial 
$b$-jets and light-quark jets, and figure of merit $S$(long-dashed line), 
for jets from decays of $Z'$ bosons with masses 2.5~TeV 
from figure~\ref{fig:eb}. 
Each multiplicity jump is considered alone:  $f_1$ (upper), $f_2$ (middle),
and $f_3$ (lower).}
\label{fig:fjsingle}
\end{figure}

The light-quark ``efficiency'' $\epsilon_q$ is the number of light-quark
jets tagged by the algorithm divided by the number of light-quark jets.
Figure~\ref{fig:fjsingle} shows the fiducial $b$-jet and light-quark efficiencies on the
basis of the inner two layers ($f_1$), the middle layers ($f_2$), and
the outer two layers ($f_3$) by themselves.  $\epsilon_b$ and $\epsilon_q$
differ significantly for thresholds $F$ above zero in all three
variables.  It is clear that the difference between the efficiencies
is larger in $f_2$ and $f_3$ than in $f_1$, a consequence of the 
double-length pixels in the innermost layer merging hits.

Figure~\ref{fig:fjsingle} also shows a significance-like figure of merit
$S\equiv\epsilon_b/2\sqrt{\epsilon_q}$ which we use to find an
optimal threshold for $B$ hadron decays in the fiducial region.
The absolute value of $S$ is unimportant; the factor
of $2$ in the denominator is for presentation purposes, so as to fit
$S$ on the same plot with the efficiencies.  In each case, $S$
rises as $F$ increases above zero, but then falls as the threshold
begins to eliminate too much signal.  The peak in $S$ is prominent in
$f_2$ and $f_3$, but less so in $f_1$, reflecting the smaller efficiency
difference in the inner two layers.  In all cases, however, a threshold
of $F=1.0$ is close to maximal $S$ while keeping high efficiency.

\begin{figure}[tb]
\centering
\includegraphics[width=.95\textwidth]{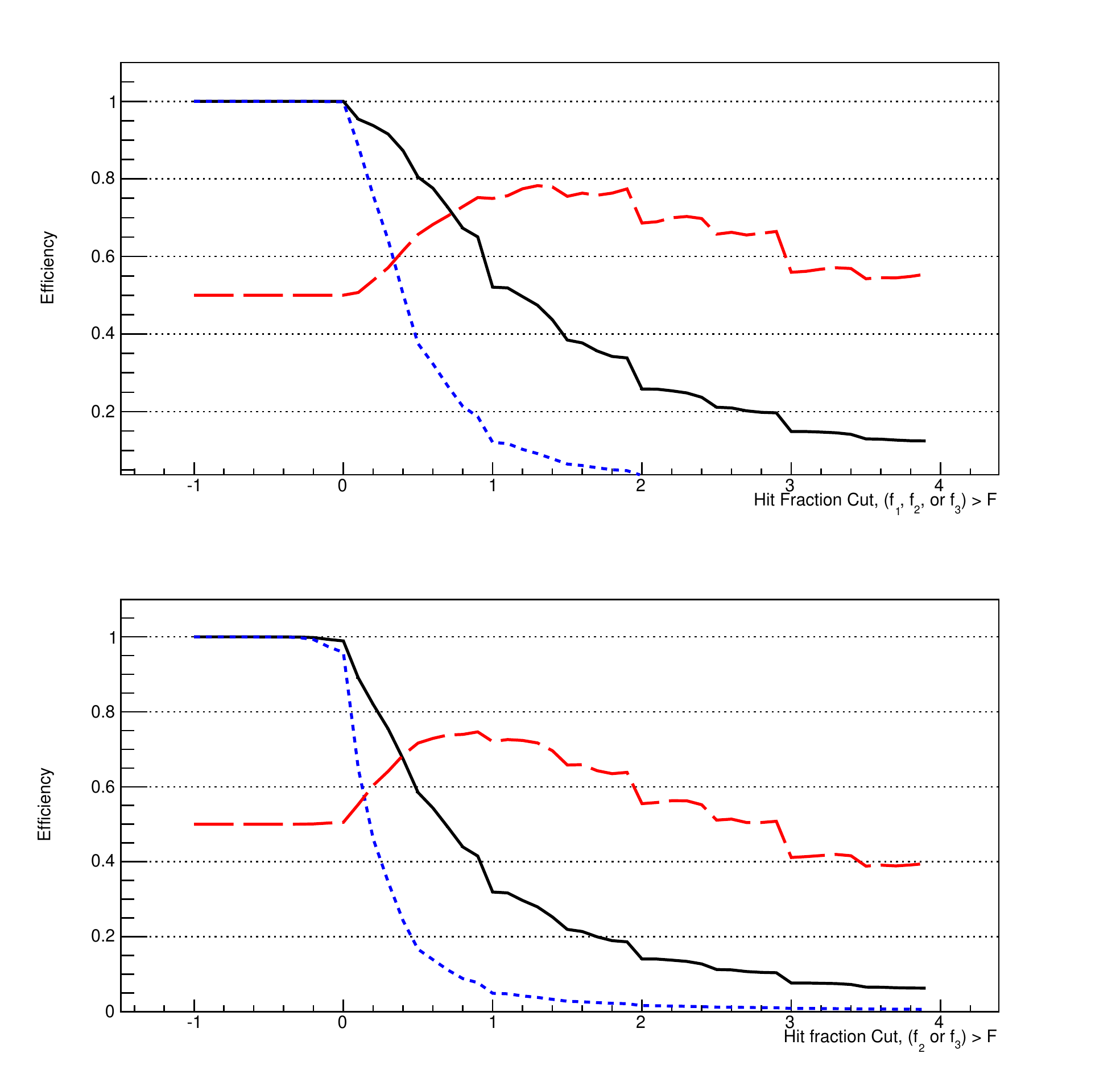}
\caption{Efficiencies for fiducial $b$-jets(solid line) 
and light-quark jets(short-dashed line),
and figure of merit $S$(long-dashed line), for jets from decays of 
$Z^{\prime}$ bosons of 2.5~TeV. 
For the upper plot, a tag is considered successful if any one 
of $f_1$, $f_2$~or $f_3 > F$ as $F$ 
runs along the horizontal axis. A successful tag on the lower plot requires $f_2$ or $f_3$. 
the fiducial region for both graphs extends from the innermost to outermost pixel layers.}
\label{fig:fjmerge}
\end{figure}

Figure~\ref{fig:fjmerge}(top) shows the efficiencies and $S$ using the whole pixel 
volume, {\em i.e.} with the fiducial region extending from the innermost to outermost
layers. A threshold of $F=1$ achieves maximal $S$. We note that because of the 
different fiducial region, $S$ in figure~\ref{fig:fjmerge} cannot be compared with 
those in figure~\ref{fig:fjsingle}. On the other hand, it is interesting to examine the 
effect of the larger pixels in layer~1: Figure~\ref{fig:fjmerge}(bottom) shows the 
efficiencies and $S$ using the whole pixel volume, but tagging only with the 
multiplicity jumps of $f_2$ and $f_3$. The figure of merit $S$ decreases only slightly,
suggesting that layer~1 adds little information overall, though maximal $S$ is achieved
at lower $F$ and thus higher efficiency.

\begin{figure}[thbp]
\centering 
%\vspace{-0.1 cm}
\includegraphics[height=.65\textwidth]{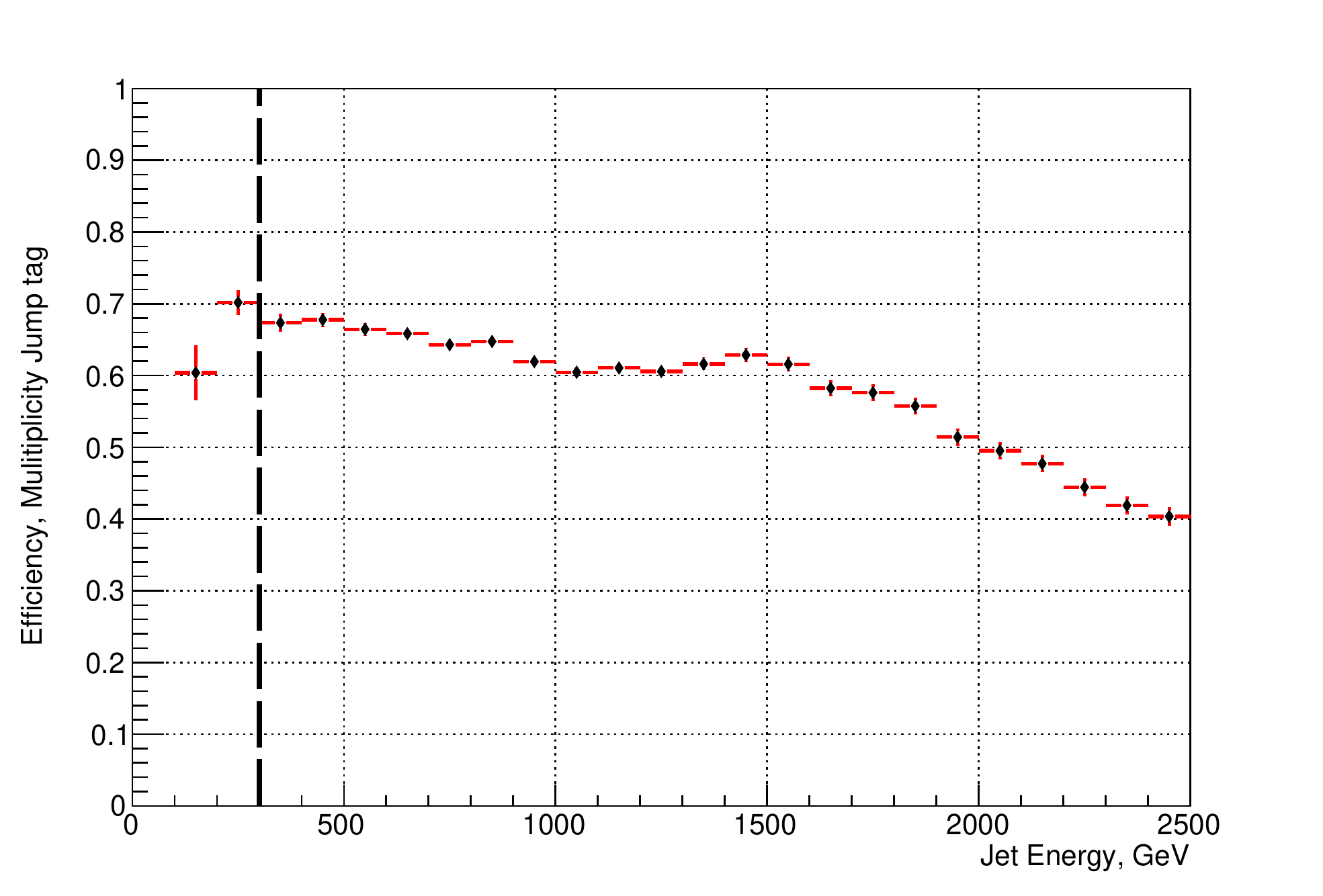}
\caption{\label{fig:HitvsE} Efficiency of multiplicity jump tagging of fiducial $b$ jets as
a function of jet energy, using all layers, full fiducial region, and threshold $F=1$. 
The dashed line indicates 600~GeV, our definition of ``extreme energy''. Both 
2.5~and 5~TeV $Z^{\prime}$ samples are used in order to improve statistics at
high energy. The percentage of light-quark jets which pass this same cut is 
also shown (open triangles).}
\end{figure}

Based upon the results presented in figures~\ref{fig:fjsingle} and~\ref{fig:fjmerge} 
we label an event as ``tagged'' when any of $f_i$ is greater than
or equal to $F =1$ 
and plot the efficiency as a function of the jet energy. This is shown in 
figure~\ref{fig:HitvsE}. We see that the efficiency indeed exhibits some of 
the expected properties. 
$\epsilon_b$ remains fairly stable even above 1~TeV.  
The figure also shows the 
percentage of uds jets which still pass the $F = 1$ cut. This fraction stays 
relatively constant with jet energy. 

Since $B$ hadrons which decay outside the outermost layer are not included in the 
fiducial volume, it is expected that the decrease in efficiency in figure~\ref{fig:HitvsE}
is due to the increasing likelihood of sequential charm decays occurring outside the 
detector volume. 

\begin{figure}[tbhp]
\centering 
\includegraphics[width=.85\textwidth]{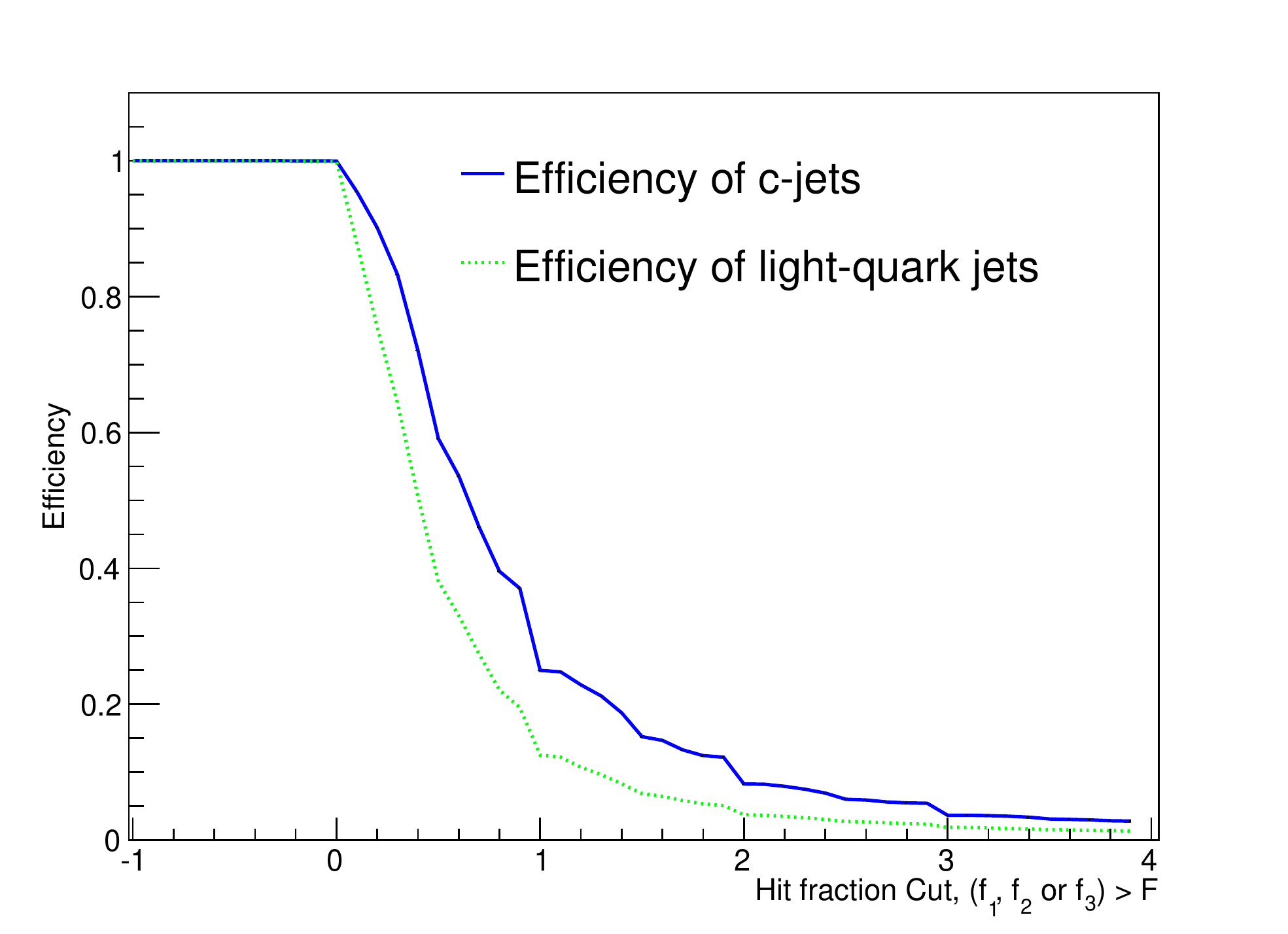}
\caption{\label{fig:HitCharm} The results of our simulation as a function of 
increasing the cut on the hit fraction difference ($f_1$, $f_2$ or $f_3\; > F$) when 
applied to jets containing a leading charm quark. The dotted curve shows jets that 
survive the cut which only contain light flavors of quarks. 
The solid curve is the efficiency for charm of the cut vs. the cut value.}
\end{figure}

It is also interesting to examine the performance of this method in distinguishing 
charm jets from light-quark jets. Figure~\ref{fig:HitCharm} shows the efficiency of 
charm versus light-quark jets.  The difference in the cut efficiency is not nearly
as pronounced as in the $b$-jet case. However, charm would still be a source of 
contamination for the multiplcity jump tagging method. 

\section{Conclusions and further study}
\label{sec:conclusion}

By examining the relative multiplicity jump $f_j$, as defined in
Equation~\ref{eq:reljump}, we have proposed an additional handle to separate
$b$ jets from those originating from light quarks.  This method does not
require charged particle tracking to function with high efficiency and accuracy
within the dense interior of extremely high energy, highly collimated jets.
Instead, simply by counting hits within a small angular region in successive
pixel layers, it maintains its efficiency to higher energies than conventional
track-based $b$ taggers.

The algorithm described in this article has intentionally been kept simple, in
part to demonstrate the feasibility of the idea by itself, but also because it
is expected that it is most likely to be used (and optimized) in combination
with other $b$ tagging techniques.  Simulation tests have already revealed that
a non-uniform pixel size when compared to the
other pixel layers may require further refinements, such as differing weights
for hits in different layers, or dynamically altering the cone used to collect 
hits based on jet energy. A neural net might be able to improve the 
discrimination power of this technique and potentially increase the difference
between the $B$ hadron efficiency and the percentage of uds jets kept. 

As a further interesting note, when the 
jet energies are as high as 4~or 5~TeV, there is a significant probability that 
the $B$ hadron will survive 
even beyond the final silicon layer used in this study, so the possibility of 
including silicon strip tracker layers, which are at even larger radii, 
within this technique is worth exploring. The prospect of 
tracking charged $B$ mesons and $B$ baryons in the detectors prior to their decay
has also not escaped our notice. 

Other complications arising from
detector geometry include overlaps between detector sensors comprising the same
layer, and the transition between cylindrical and endcap disk layers.  Effects
not included in the simulation include pileup, {\it i.e.}, multiple
interactions in the same beam crossing, and potential hadronic interactions
between $B$ hadrons and the material it traverses.  In spite of these
simplifications, however, this study suggests that a relative multiplicity jump
is a promising observable to improve $b$ tagging at the extreme energies
increasingly required to probe for new physics at the energy frontier.

If shown to work in the LHC detectors this technique could have implications for the detector 
design at future colliders such as the Future Circular Collider (FCC)\cite{FCC}. 
Such a machine would produce jets with a 5~TeV $B$~hadron. 
Extending finely segmented pixel coverage to larger radii 
in order to tag these jets may be desirable for such future detectors.

\section*{Acknowledgments}

We thank Juan Rojo, Cigdem Issever, and Anthony Weidberg 
for their critical thoughts and advice prior to publication. We additionally 
thank Juan Rojo for his advice on theoretical models. 
This work was supported by the Science and Technology Facilities 
Council of the United Kingdom grant number ST/N000447/1 
and the Higher Education Funding Council of England.

\section*{References}

\bibliography{JumpMultbibTCJ2p1}{}

\providecommand{\href}[2]{#2}\begingroup\raggedright\begin{thebibliography}{10}

\bibitem{Behr:2015oqq}
J.~K. Behr, D.~Bortoletto, J.~A. Frost, N.~P. Hartland, C.~Issever and J.~Rojo,
  \emph{{Boosting Higgs pair production in the $b\bar{b}b\bar{b}$ final state
  with multivariate techniques}},  \href{http://arxiv.org/abs/1512.08928}{{\tt
  1512.08928}}.

\bibitem{Gouzevitch:2013qca}
M.~Gouzevitch, A.~Oliveira, J.~Rojo, R.~Rosenfeld, G.~P. Salam and V.~Sanz,
  \emph{{Scale-invariant resonance tagging in multijet events and new physics
  in Higgs pair production}},
  \href{http://dx.doi.org/10.1007/JHEP07(2013)148}{\emph{JHEP} {\bf 07} (2013)
  148}, [\href{http://arxiv.org/abs/1303.6636}{{\tt 1303.6636}}].

\bibitem{Alwall:2008ag}
J.~Alwall, P.~Schuster and N.~Toro, \emph{{Simplified Models for a First
  Characterization of New Physics at the LHC}},
  \href{http://dx.doi.org/10.1103/PhysRevD.79.075020}{\emph{Phys. Rev.} {\bf
  D79} (2009) 075020}, [\href{http://arxiv.org/abs/0810.3921}{{\tt
  0810.3921}}].

\bibitem{ATL-PHYS-PUB-2015-022}
{\scshape ATLAS} collaboration, \emph{{Expected performance of the ATLAS
  $b$-tagging algorithms in Run-2}},  Tech. Rep. ATL-PHYS-PUB-2015-022, CERN,
  Geneva, Jul, 2015.

\bibitem{Aaboud:2016nbq}
{\scshape ATLAS} collaboration, M.~Aaboud et~al., \emph{{Search for resonances
  in the mass distribution of jet pairs with one or two jets identified as
  $b$-jets in proton--proton collisions at $\sqrt{s}=13$ TeV with the ATLAS
  detector}},
  \href{http://dx.doi.org/10.1016/j.physletb.2016.05.064}{\emph{Phys. Lett.}
  {\bf B759} (2016) 229--246}, [\href{http://arxiv.org/abs/1603.08791}{{\tt
  1603.08791}}].

\bibitem{Pedersen:2015knf}
K.~Pedersen and Z.~Sullivan, \emph{{$\mu_x$ boosted-bottom-jet tagging and Z′
  boson searches}},
  \href{http://dx.doi.org/10.1103/PhysRevD.93.014014}{\emph{Phys. Rev.} {\bf
  D93} (2016) 014014}, [\href{http://arxiv.org/abs/1511.05990}{{\tt
  1511.05990}}].

\bibitem{deFavereau:2013fsa}
{\scshape DELPHES 3} collaboration, J.~de~Favereau, C.~Delaere, P.~Demin,
  A.~Giammanco, V.~Lemaître, A.~Mertens et~al., \emph{{DELPHES 3, A modular
  framework for fast simulation of a generic collider experiment}},
  \href{http://dx.doi.org/10.1007/JHEP02(2014)057}{\emph{JHEP} {\bf 02} (2014)
  057}, [\href{http://arxiv.org/abs/1307.6346}{{\tt 1307.6346}}].

\bibitem{AtlasMaterial}
G.~Aad et~al., \emph{{ATLAS pixel detector electronics and sensors}},
  \href{http://dx.doi.org/10.1088/1748-0221/3/07/P07007}{\emph{JINST} {\bf 3}
  (2008) P07007}.

\bibitem{JumpMult}
A.~M. Halling and S.~Kwan, \emph{{A Multiplicity jump trigger for fixed target
  charm and beauty experiments}},
  \href{http://dx.doi.org/10.1016/0168-9002(93)91174-L}{\emph{Nucl. Instrum.
  Meth.} {\bf A333} (1993) 324--329}.

\bibitem{Agostinelli:2002hh}
{\scshape GEANT4} collaboration, S.~Agostinelli et~al., \emph{{GEANT4: A
  Simulation toolkit}},
  \href{http://dx.doi.org/10.1016/S0168-9002(03)01368-8}{\emph{Nucl. Instrum.
  Meth.} {\bf A506} (2003) 250--303}.

\bibitem{Allison:2006ve}
J.~Allison et~al., \emph{{Geant4 developments and applications}},
  \href{http://dx.doi.org/10.1109/TNS.2006.869826}{\emph{IEEE Trans. Nucl.
  Sci.} {\bf 53} (2006) 270}.

\bibitem{Pythia8}
T.~Sjostrand, S.~Mrenna and P.~Z. Skands, \emph{{A Brief Introduction to PYTHIA
  8.1}}, \href{http://dx.doi.org/10.1016/j.cpc.2008.01.036}{\emph{Comput. Phys.
  Commun.} {\bf 178} (2008) 852--867},
  [\href{http://arxiv.org/abs/0710.3820}{{\tt 0710.3820}}].

\bibitem{Skands:2014pea}
P.~Skands, S.~Carrazza and J.~Rojo, \emph{{Tuning PYTHIA 8.1: the Monash 2013
  Tune}}, \href{http://dx.doi.org/10.1140/epjc/s10052-014-3024-y}{\emph{Eur.
  Phys. J.} {\bf C74} (2014) 3024}, [\href{http://arxiv.org/abs/1404.5630}{{\tt
  1404.5630}}].

\bibitem{Peterson:1982ak}
C.~Peterson, D.~Schlatter, I.~Schmitt and P.~M. Zerwas, \emph{{Scaling
  Violations in Inclusive e+ e- Annihilation Spectra}},
  \href{http://dx.doi.org/10.1103/PhysRevD.27.105}{\emph{Phys. Rev.} {\bf D27}
  (1983) 105}.

\bibitem{Lange:2001uf}
D.~J. Lange, \emph{{The EvtGen particle decay simulation package}},
  \href{http://dx.doi.org/10.1016/S0168-9002(01)00089-4}{\emph{Nucl. Instrum.
  Meth.} {\bf A462} (2001) 152--155}.

\bibitem{AtlasTDR}
M.~Capeans, G.~Darbo, K.~Einsweiller, M.~Elsing, T.~Flick, M.~Garcia-Sciveres
  et~al., \emph{{ATLAS Insertable B-Layer Technical Design Report}},  Tech.
  Rep. CERN-LHCC-2010-013. ATLAS-TDR-19, CERN, Geneva, Sep, 2010.

\bibitem{FastJet}
M.~Cacciari, G.~P. Salam and G.~Soyez, \emph{{FastJet User Manual}},
  \href{http://dx.doi.org/10.1140/epjc/s10052-012-1896-2}{\emph{Eur. Phys. J.}
  {\bf C72} (2012) 1896}, [\href{http://arxiv.org/abs/1111.6097}{{\tt
  1111.6097}}].

\bibitem{antikt}
M.~Cacciari, G.~P. Salam and G.~Soyez, \emph{{The Anti-k(t) jet clustering
  algorithm}},
  \href{http://dx.doi.org/10.1088/1126-6708/2008/04/063}{\emph{JHEP} {\bf 04}
  (2008) 063}, [\href{http://arxiv.org/abs/0802.1189}{{\tt 0802.1189}}].

\bibitem{FCC}
{\scshape TLEP Design Study Working Group} collaboration, M.~Bicer et~al.,
  \emph{{First Look at the Physics Case of TLEP}},
  \href{http://dx.doi.org/10.1007/JHEP01(2014)164}{\emph{JHEP} {\bf 01} (2014)
  164}, [\href{http://arxiv.org/abs/1308.6176}{{\tt 1308.6176}}].

\end{thebibliography}\endgroup
\bibliographystyle{JHEP}

\end{document}